\documentclass[10,A4,conference]{IEEEtran}
\usepackage {algorithm}
\usepackage {algpseudocode}
\usepackage{amsthm,amssymb,graphicx,multirow,amsmath,color,amsfonts}
\usepackage[update,prepend]{epstopdf}
\usepackage[noadjust]{cite}
\usepackage[latin1]{inputenc}
\usepackage{tikz}
\usepackage{bbm} 
\usepackage{pdfpages}
\usepackage{multirow}
\usepackage{comment}

\usepackage{algorithm}
\newtheorem{definition}{Definition}
\newtheorem{lemma}{Lemma}
\usepackage{balance}

\usepackage{optidef}

\include{notation}
\allowdisplaybreaks 

\usepackage{setspace}	

\begin{document}
\graphicspath{{./Figures/}}
\title{
Multi-sided Matching For Space-Air-Ground Integrated Systems
}

\author{
  \IEEEauthorblockN{
    Abdoul Karim A. H. Saliah\IEEEauthorrefmark{2},Doha Hamza\IEEEauthorrefmark{5},
    Hajar El Hammouti\IEEEauthorrefmark{2},
    Jeff S. Shamma\IEEEauthorrefmark{1}\IEEEauthorrefmark{4},
    Mohamed-Slim Alouini\IEEEauthorrefmark{1}
  }

    \IEEEauthorblockA{
    \IEEEauthorrefmark{2} College of Computing, Mohammed VI Polytechnic University (UM6P), Benguerir, Morocco.
  }

     \IEEEauthorblockA{
    \IEEEauthorrefmark{5} Electrical Engineering Department, University of Prince Mugrin (UPM), Madinah, KSA.
  }

  \IEEEauthorblockA{
    \IEEEauthorrefmark{4} Industrial and Enterprise Systems Engineering (ISE), Urbana-Champaign, Chicago, USA.
  }

 \IEEEauthorblockA{
    \IEEEauthorrefmark{1} CEMSE division, King Abdullah University of Science and Technology (KAUST), Thuwal, Makkah Province, KSA.
  }
 \IEEEauthorblockA{
      Email: \{abdoul.saliah,hajar.elhammouti\}@um6p.ma, d.hamza@upm.edu.sa, \{jeff.shamma, mohamed.alouini\}@kaust.edu.sa
}
}

\maketitle

\begin{abstract}

Space-air-ground integrated networks (SAGINs) will play a pivotal role in 6G communication systems. They are considered a promising technology for enhancing network capacity in densely populated urban areas and extending connectivity to rural regions. However, the complex, multi-layered, and heterogeneous nature of SAGINs demands an innovative approach to designing their multi-tier associations. In this context, we propose a modeling of the SAGINs association problem using multi-sided matching theory. Our objective is to devise a reliable, asynchronous, and fully distributed approach that associates nodes across the layers to maximize the total end-to-end rate of the assigned agents. To achieve this, our problem is formulated as a multi-sided many-to-one matching game. We introduce a randomized matching algorithm with minimal information exchange. The algorithm is shown to reach an efficient and stable association between nodes in adjacent layers. Simulation results show that our proposed approach yields significant gains compared to both greedy and distance-based algorithms.

\end{abstract}


\section{Introduction} \label{sec:intro}


Space-air-ground integrated networks (SAGINs) have emerged as a promising solution for ensuring ubiquitous and cost-effective connectivity globally. In their design, flying infrastructure such as satellites, high altitude platforms (HAPs), and unmanned aerial vehicles (UAVs) collaborate with the terrestrial network to form a coordinated system. This integration aims to deliver high network connectivity, particularly in remote and inaccessible areas~\cite{zhang2023HAP,ndiaye2022age}. SAGINs are envisioned to provide support for damaged terrestrial infrastructure after natural disasters and to extend network coverage during emergencies and temporary events. Additionally, they can offer backhaul resources, helping to offload traffic from terrestrial networks. SAGINs also exhibit superior energy efficiency compared to traditional terrestrial networks. Satellites and HAPs, in particular, benefit from consistent solar radiation, which ensures a constant supply of green energy.

Nevertheless, the deployment of SAGINs poses several technical challenges including resource allocation, network management, and mobility~\cite{liu2018space}. While these problems were addressed for standalone spacial/aerial networks and terrestrial systems, there are still many unanswered questions in combined multi-layered systems. First, compared to traditional networks, SAGINs serve a larger number of users, resulting in high user dynamics. This necessitates the development of fast convergence algorithms to ensure rapid resource allocation~\cite{jia2022SAGIN}. Second, due to the heterogeneous nature of SAGINs, it is important to provide customized resource allocation approaches that adapt to the multi-layered network structure of SAGINs. Consequently, fully distributed mechanisms that circumvent the need for extensive coordination and computational resources are highly desirable.
 Over the past decade, numerous works studied resource allocation for aerial and terrestrial networks both separately, as standalone networks, and jointly as an integrated system~\cite{el2021optimal}. Previous works essentially proposed centralized approaches. In general, the multi-layer optimization is decoupled into sub-problems, each sub-problem is tackled separately using a specific centralized optimization technique~\cite{alsharoa2020improvement,abderrahim2020latency,shi2020joint}. However, while centralized approaches efficiently solve multi-variate optimization problems, they are not flexible enough to easily adapt to a network of multiple communication layers. Furthermore, they require a centralized entity that coordinates the actions of agents. Such approaches also involve a large amount of overhead and information exchange and require global knowledge of the network \cite{arabi2019rat}. 

In contrast to existing works, our paper presents a distributed, asynchronous, and computationally efficient algorithm that achieves a well-balanced association across the multi-tier network while satisfying the network constraints. We find game-theoretic modeling, particularly matching theory, to be a compelling framework to address the studied problem. Matching theory started with the work by Gale and Shapley~\cite{gale1962college} and quickly became an important resource allocation tool, initially for two-sided markets and later for their extensions~\cite{roth1992two}. There are two main categories within matching theory: marriage markets and transferrable utility games (TU), also called assignment games. The outcome of a TU game is not only the matching but also payoff vectors to the agents that add up to the value that can be generated between the pair of users \cite{shapley1971assignment}. We focus on TU in our model. 

One algorithm that serves our purpose of having a decentralized mechanism for resource allocation is the Blind Matching Algorithm (BLMA) in~\cite{hamza2017blma}. The authors propose a randomized matching algorithm characterized by probabilistic activation of agents in a two-sided market. Agents have aspiration levels, utilities they currently have or have accrued in the past, that tell them whether a match with another active agent from the other side of the market is better than their current state. This is the only information the agents use to make their decisions. Despite this knowledge limitation, the algorithm is shown to reach an equilibrium. We use the idea of these blind and random encounters between agents of adjacent layers to associate the agents of the multi-layered SAGIN.

\begin{figure}
\centering
\includegraphics[scale=.28]{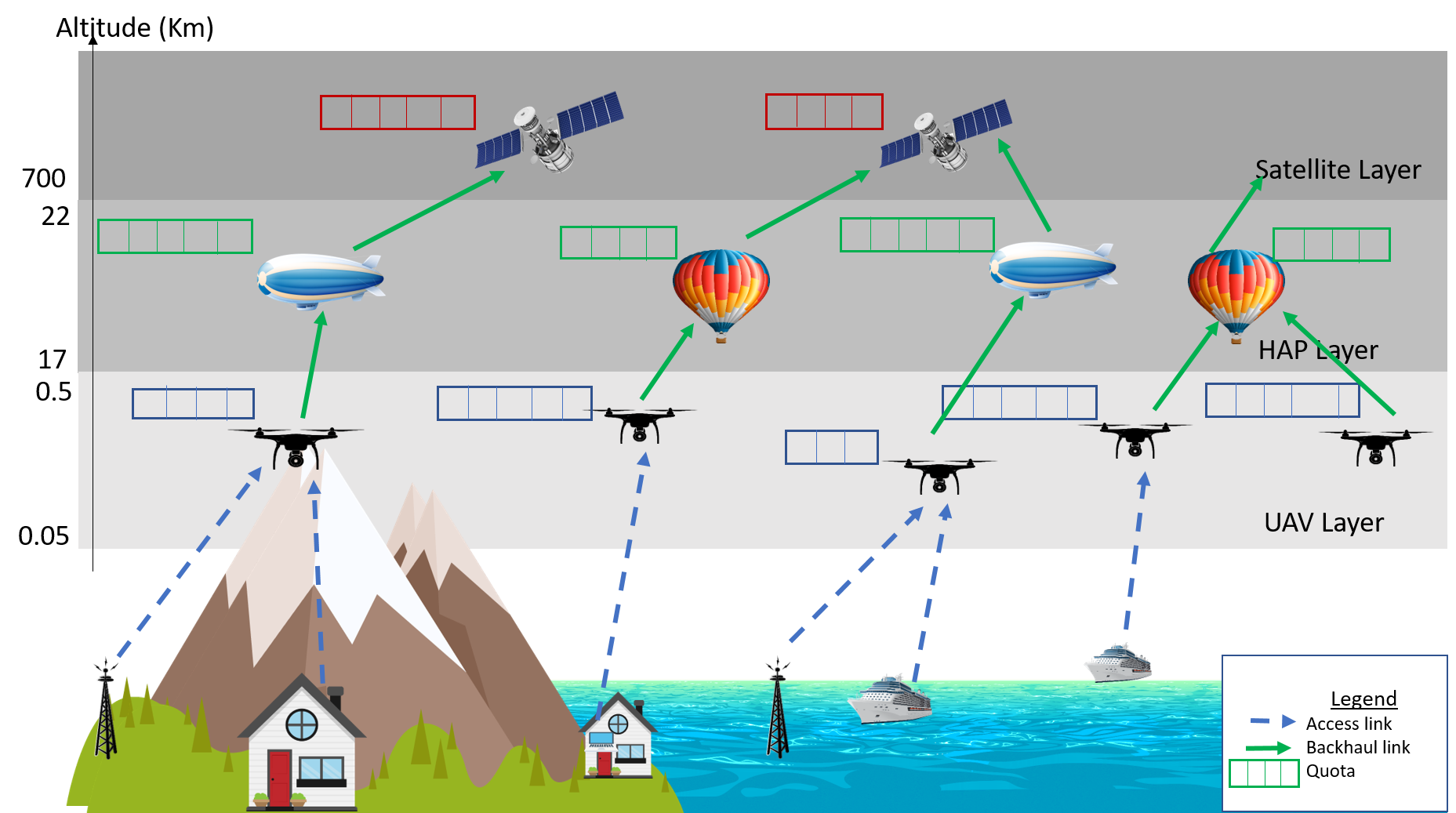}
\caption{The studied system model}
\label{fig:my_label}
\end{figure}

There are two important solution concepts in matching theory that define equilibrium. \textit{Pairwise stability} ensures that two agents cannot form a blocking pair to a matching. On the other hand, the \textit{core} defines the set of feasible payoff vectors that cannot be improved upon by any coalition of players \cite{shapley1971assignment}. These two concepts align in several matching markets. Although \cite{hamza2017blma} presents an algorithm to achieve a pairwise stable state that is also core-stable for one-to-one matching markets, extending this to the multi-sided case is generally challenging, as the core may be empty. Nonetheless, works like \cite{atay2017essays} suggest that meticulously designing the value generated by a tuple of agents can ensure the non-emptiness of the core.
 

In this context, we formulate the SAGINs association problem as a multi-sided many-to-one matching market. We design the value generated by a coalition of users in the game to guarantee the non-emptiness of the core. Subsequently, we propose a blind and randomized matching algorithm to form groups of agents from different layers, maximizing the overall utility of the network represented by the end-to-end rate.

Although some works exist in wireless communications on multi-sided matching, it is generally restricted to three-sided marriage markets, e.g., \cite{zhao2018resource, liu2019joint}. To the best of the authors' knowledge, this is the first work within wireless communications that suggests the use of a multi-sided assignment market with a TU framework, for the SAGINs association problem.


In the remainder of the paper, we describe the studied system model. Then, we provide details about the proposed approach and exhibit its advantages. Next, we show the performance of the blind multi-sided matching algorithm compared to the greedy algorithm and distance-based association. Finally, we provide some concluding remarks.


\section{System Model}

\subsection{Communication Model}

We assume an uplink communication where UAVs and HAPs act as relays that retransmit data traffic to the satellites. Each UAV communicates both with the users and HAPs through the air-to-ground and air-to-air channels. Similarly, HAPs interact with UAVs and satellites via air-to-air and air-to-space channels.  For communication with the space tier, a user initially sends its data to a UAV, which then forwards the traffic to one of the HAPs. The HAP, in turn, transfers the traffic to a selected satellite. An illustration of the studied system model is provided in Figure~\ref{fig:my_label}.

Let $\mathcal{S}^k$ be the set of nodes in layer $\{0,1,2,3\}$. Here, $\mathcal{S}^0$ denotes the set of devices on the ground, $\mathcal{S}^1$ the set of UAVs, $\mathcal{S}^2$ the set of HAPs, and $\mathcal{S}^3$ the set of satellites. Let $\mathcal{S}= \cup_{k=0}^{3}\mathcal{S}^k$ be the set of all nodes.
 The ground-to-air channel between a user $i$ and a UAV $j$ is expressed as \cite{hammouti2018air}
\begin{equation}
L_{ij}^{\rm G2A}\!=\! 20 \log\! \left(\!\frac{4 \pi f_c d_{ij}^{\rm G2A}}{c}\!\!\right) \!+ \!P^{\rm LOS}_{ij} \eta_{\rm LOS}^{\rm G2A} + \left(1-P^{\rm LOS}_{ij}\right) \eta_{\rm NLOS}^{\rm G2A},  
\end{equation}

\noindent where $f_c$ is the carrier frequency in Hz, $c$ is the velocity of light in $\mathrm{m} / \mathrm{s}$, $\eta_{\rm LOS}^{\rm G2A}$ and $\eta_{\rm NLOS}^{\rm G2A}$ are environment-dependent parameters that denote the average additive losses caused by the free space path-loss for line of sight (LOS) and non-LOS (NLOS) links in the ground-to-air channel propagation in $\mathrm{dB}$, respectively. Also, $d_{ij}^{\rm G2A}$ is the distance between the user $i$ and UAV $j$, $P^{\rm LOS}_{ij}$ is the LOS probability of the user $i$ to UAV $j$ link, which can be calculated as
\begin{equation}
P^{\rm LOS}_{ij}=\frac{1}{1+\phi \exp \left(-\varphi\left(\theta_{ij}-\phi\right)\right)},
\end{equation}
where $\phi$ and $\varphi$ are constant values determined by the environment, $\theta_{ij}$ is the elevation angle between user $i$ and UAV $j$, which is expressed as
\begin{equation}
\theta_{ij}=\arctan \left(\frac{h_j}{\sqrt{(d_{ij}^{\rm G2A})^2-h_j^2}}\right),
\end{equation}
where $h_j$ is the altitude of UAV $j$. 

The channels between UAVs and HAPs, and between HAPs and satellites, are modeled as a LOS link and expressed as

\begin{equation}
L_{ij}^{\rm A2A}=20 \log \left(\frac{4 \pi f_c d_{ij}^{\rm A2A}}{c}\right)+\eta_{L O S}^{\rm A2A},
\end{equation}

\noindent with $d_{ij}^{\rm A2A}$ the distance between UAV $i$ and HAP $j$ or HAP $i$ and satellite $j$, and $\eta_{\rm LOS}^{\rm A2A}$ is the average additive loss caused by the free space path-loss for LOS in the air-to-air channel propagation.




In order to avoid interference, we apply Frequency-Division Multiplexing (FDM) to the network. Accordingly, the data rate between two nodes $i\in \mathcal{S}^k$ and $j\in \mathcal{S}^{k+1}$, $k \in \{0,1,2\}$, in adjacent layers is given by
\begin{equation}
r_{ij}^{k,k+1}=B_{ij}\log _2\left(1+\frac{P_ig_{ij}^{k,k+1}}{N_0 B_{ij}}\right),
\end{equation}
where $B_{ij}$ is the allocated bandwidth between nodes $i$ and $j$, $g_{ij}^{0,1}=10^{-L_{ij}^{\rm G2A}/10}$, and $g_{ij}^{k,k+1}=10^{-L_{ij}^{\rm A2A}/ 10}$ for $k \in \{1,2\}$, $P_i$ is the transmit power of node $i$ and $\sigma^2$ is the noise power spectral density.  Moreover, for convenience, we consider that the network performs during a small interval of time during which the network configuration remains fixed.



\subsection{Problem Formulation}
 
 Our objective is to maximize the end-to-end data rates of the ground users. The end-to-end data rate from a ground user to the space tier is given by the minimum rate of the relaying links. Indeed, the performance of the communication between the ground tier and the space tier depends on the link that experiences the lowest data rate performance. This link represents the bottleneck in the relaying system, thereby impacting the overall performance. In this context, we need to select which associations are to be made among adjacent layers for all agents in the network such that the objective function is maximized. Let $a_{ij}^{k,k+1}$ be a binary variable associated with a link $(i,j)$, where node $i$ belongs to layer $k$ and $j$ belongs to $k+1$, i.e.,
$$
a_{ij}^{k,k+1}=\begin{cases}
    1& \text{if } (i,j) \in \mathcal{S}^k\times  \mathcal{S}^{k+1 }\text{ is active}\\
    0& \text{otherwise.}\\
\end{cases}
$$
 
 We assume that each node in $\mathcal{S}^k$ can be associated with at most one node in $\mathcal{S}^{k+1}$, while a node $j$ in $\mathcal{S}^{k+1}$ can serve a group of downstream nodes with a maximum quota $q_j^{k+1}$. These constraints are expressed as follows
 \begin{eqnarray}
    && \sum_{i\in{\cal S}^k} {a_{ij}^{k,k+1}} \leq q_j^{k+1}, \forall k\in\{0,1,2\}, \forall j\in{\cal S}^{k+1}, \label{equ1}\\
    && \sum_{j\in{\cal S}^{k+1}} {a_{ij}^{k,k+1}} \leq 1, \forall k\in\{0,1,2\}, \forall i\in{\cal S}^{k}, \label{equ2}
 \end{eqnarray}

Accordingly, the end-to-end rate optimization is constrained by $(\ref{equ1})$ and $(\ref{equ2})$. Let $\boldsymbol{a}$ be the matrix of associations. Our end-to-end data rate maximization is written as follows
 


\begin{maxi!}
{\boldsymbol{a}}{\! \sum_{(\!i,j,l,m\!) \in \mathcal{S}^0\!\times \!\!\mathcal{S}^1 \!\!\times \mathcal{S}^2\!\! \times\! \!\mathcal{S}^3} \!\!\!\!\!\!a_{ij}^{0,1}a_{jm}^{1,2}a_{ml}^{2,3}min(r_{ij} ^{0,1},r_{jm} ^{1,2}, r_{ml} ^{2,3})\label{objective}}
{\label{GeneralOptimizationOrigin}}{}
\addConstraint{ (\ref{equ1})\text{ and }(\ref{equ2})\label{temps}}
{}{}
\addConstraint{ a_{ij}^{k,k+1}~\!\!\!\!\in \!\!\{0\!,\!1\},  \forall (i,j)\! \in\! \mathcal{S}^{k}\!\!\times\!\! \mathcal{S}^{k+1}\!\!, \forall k\!\in \!\{\!0,\!1,\!2\!\}  \label{accuracy}}{}{}
\end{maxi!}

 Due to the integer nature of the association variable, i.e., a binary variable, and the quota constraint, the underlying optimization problem is a challenging one. To circumvent this difficulty, we formulate our optimization problem within the framework of matching theory.


\section{A Multi-sided Assignment Game}
\label{Approach}
In this section, we begin by introducing some definitions for coalitional games and core allocations. Then we formulate the studied problem as a multi-sided assignment game. 

\subsection{Defining Coalitional and Multisided Assignment Games}

Cooperative game theory captures the cooperation among interacting agents. Within this framework, one prevalent category of games is called \textit{coalitional games with transferable utility} (CGTU). CGTU are defined over a set of players $\mathcal{N}$ and are determined by a the coalitional function $v: 2^{\mathcal{N}} \rightarrow \mathbb{R}$, also called characteristic function, such that, for each non-empty set $\mathcal{C} \subseteq \mathcal{N}$, also called a \textit{coalition}, the value $v(\mathcal{C})$ represents the collective gain achievable by the members of $\mathcal{C}$ through cooperation. In the following, we provide a mathematical definition of CGTU.

\begin{definition}[Coalitional game with transferable utility (CGTU)]
Let $\mathcal{N} \subset \mathbbm{N}$, and $v: 2^\mathcal{N}\xrightarrow{} \mathbbm{R}$ a function such that $v(\emptyset)=0$. The pair $(\mathcal{N},v)$ is called a \textit{transferable utility coalitional game}, with $v$ its coalitional function.
\end{definition}

In CGTU, it is assumed that players among the coalition can distribute the gain among themselves to achieve a desirable stable outcome. This outcome assigns a payoff to each agent in $\mathcal{N}$, and is captured by a payoff vector $\mathbf{x}=(x_1,\dots,x_N) \in \mathbb{R}^N$. The main goal of these games is to identify favorable payoff vectors $\mathbf{x}$ that guarantee properties such as fairness and stability among the players. These types of allocation payoffs are referred to as solution concepts. A well-established solution concept used in the literature to address the stability challenge in these games is the \textit{core}. The core defines the set of all allocations that demonstrate stability and ensure that no coalition has an incentive to deviate. The following definition presents a mathematical formulation of the core.

\begin{definition}[Core]
 Suppose a CGTU $(\mathcal{N},v)$. The core of the game $(\mathcal{N},v)$ is the set of payoff vectors $\mathbf{x}=(x_1,\dots,x_N)\in \mathbbm{R}^N$ such that,
 \begin{equation}
     \forall \mathcal{C} \subset \mathcal{N}, \sum \limits_{i \in \mathcal{C}}x_i\geq v(\mathcal{C}).
 \end{equation}

\end{definition}


\subsection{$K$-sided assignment game for SAGIN association}
In this context, a multi-sided assignment game with $K$ sets of players can be viewed as a CGTU, where a coalition can be formed only if it contains exactly one agent from each of the $K$ sectors.

 A multi-sided assignment game is defined with respect to a \textit{multi-sided assignment market}. A multi-sided assignment market $\gamma=(\mathcal{N}_0,\mathcal{N}_1,...,\mathcal{N}_{K-1}; A)$, is characterized by $K$ sets of players denoted as $\mathcal{N}_0, \dots, \mathcal{N}_{K-1}$, and an assignment mapping ${A}: \mathcal{N}_0 \times \dots \times \mathcal{N}_{K-1} \rightarrow \mathbbm{R}$, where $A_{\mathcal{C}_0}$ represents the value created by the coalition $\mathcal{C}_0=(i_0, i_2, \dots, i_{K-1}) \in \mathcal{N}_0 \times \dots \times \mathcal{N}_{K-1}$ called basic coalition. We will use the notation $\mathcal{C}_0$ to refer to a basic coalition in the rest of the paper. In the following definition, we provide a mathematical formulation of a multi-sided assignment game.

\begin{definition}(Multi-sided assignment game). The multi-sided assignment game corresponding to the multi-sided assignment market  $\gamma=(\mathcal{N}_0,\mathcal{N}_2,...,\mathcal{N}_{K-1}; A)$ is a pair $(\mathcal{N},\upsilon_A)$ where $\mathcal{N}=\cup_{k=0}^{K-1}N_k$ is the set of players and $\upsilon_A$, the characteristic function, is defined as
\begin{equation}
  \upsilon_A(\mathcal{C})=\max_{\mu\in {\cal M}_\mathcal{C}}\sum_{{\mathcal{C}_0}\in \mu}{A_{\mathcal{C}_0}},\forall \mathcal{C}\subseteq \mathcal{N},   
\end{equation}

\noindent where ${\cal M}_\mathcal{C}={\cal M}(\mathcal{C}\cap \mathcal{N}_0,\mathcal{C}\cap \mathcal{N}_1,...,\mathcal{C}\cap \mathcal{N}_K)$ is the set of all matchings for a coalition $\mathcal{C}\subseteq \mathcal{N}$.
A matching for a coalition $\mathcal{C}\subseteq N$ is a subset of basic coalitions such that each player belongs to at most one basic coalition. 
\end{definition}
As a CGTU, it is relevant to assess the existence of the core and identify it for a $K$-sided assignment game. In the next subsection, we explain how the studied SAGIN association problem can be modeled as a $K$-sided assignment game. Then, we show the non-emptiness of the core for this game.  
\subsection{SAGIN as a Multi-sided Assignment Game}


In the SAGIN association problem, our objective is to determine the associations among adjacent layers for all agents in the network, with the aim of maximizing the end-to-end rate. 
In the framework of a $K$-sided assignment game, a coalition can be formed only if it includes exactly one agent from each of the $K$ sectors.


In this context, we consider the value of a basic coalition as the sum of the values of its building layers: the user-UAV valuation, plus the UAV-HAP valuation, and the HAP-satellite valuation. 
In the following, we define the 4-sided SAGIN assignment game, which verifies the local additivity. 

\begin{definition}[4-sided SAGIN assignment game]
The $4$-sided SAGIN assignment game $(\mathcal{S},\upsilon_A)$ is associated to a market.
The $4$-sided SAGIN assignment market $({\cal S}^0,\mathcal{S}^1,\mathcal{S}^2,\mathcal{S}^3;A)$ satisfies local additivity, so there exists a set of matrices $\mathbf{b}^{k,k+1}$ for $k\in\{0,1,2\}$ with

\begin{equation}
    \mathbf{b}^{k,k+1} =(\frac{r_{ij}^{k,k+1}}{q_j})_{(i,j)\in {\cal S}^k\times {\cal S}^{k+1}}
 \end{equation}

such that the expression of the valuation matrix is:
\begin{equation}
\begin{aligned}
    A_{{\mathcal{C}_0}} &= \mathbf{b}^{0,1}(i_0,i_1) + \mathbf{b}^{1,2}(i_1,i_2) + \mathbf{b}^{2,3}(i_2,i_3), \\
    &\quad \forall {\mathcal{C}_0}=(i_0,i_1,i_2,i_3) \in \prod_{k=0}^3{{\cal S}}^k.
\end{aligned}
\end{equation}
 

\end{definition}


Local additivity is a useful property in our market. The following lemma is due to \cite{stuart1997supplier}.

\begin{lemma}\label{lemma}
Let $({\cal S}^0,{\cal S}^1,{\cal S}^2,{\cal S}^3;A)$ be a locally additive multi-sided market. The core of the corresponding $K$-sided assignment game $(\mathcal{S},\upsilon_A)$ is non-empty.
\end{lemma}

Considering the local additivity of the 4-sided SAGIN assignment market, Lemma \ref{lemma} guarantees the non-emptiness of the core of our game $(\mathcal{S},\upsilon_A)$. Due to the chain structure of the market, we can equally focus on the adjacent layers' valuation matrices instead of the full $K$-dimensional matrix~\cite{stuart1997supplier}.

Within the framework of multi-sided assignment games, markets with many sectors introduce challenges to defining stability. We refer the interested reader to \cite{atay2018bargaining, atay2017essays} for more details on the topic. We are particularly interested in a specific notion of stability defined between two agents on opposite side of the market for two given layers. A matching between two adjacent layers is $\epsilon$-stable if all matched agents attain their value and no $\epsilon$-improvement to two agents' utility is possible~\cite{hamza2017blma}. $\epsilon$-stability introduces the concept of $\epsilon$-core, a set of payoffs formed by relaxing the  traditional core constraints. When the non-emptiness of the game's core is confirmed, $\epsilon$-core converges to the core as $\epsilon$ approaches zero.

Our modeling thus far proved fruitful. We formulated the SAGIN problem as a multi-sided assignment game with a non-empty core. We have an idea of what it means to have an optimal matching, but a question remains as to how to achieve it? We address this next by proposing an algorithm to maximize the sum of the end-to-end rate and achieve equilibrium.

\section{A Blind multi-sided matching algorithm}
\begin{figure*}
    \centering
    \includegraphics[scale=0.5]{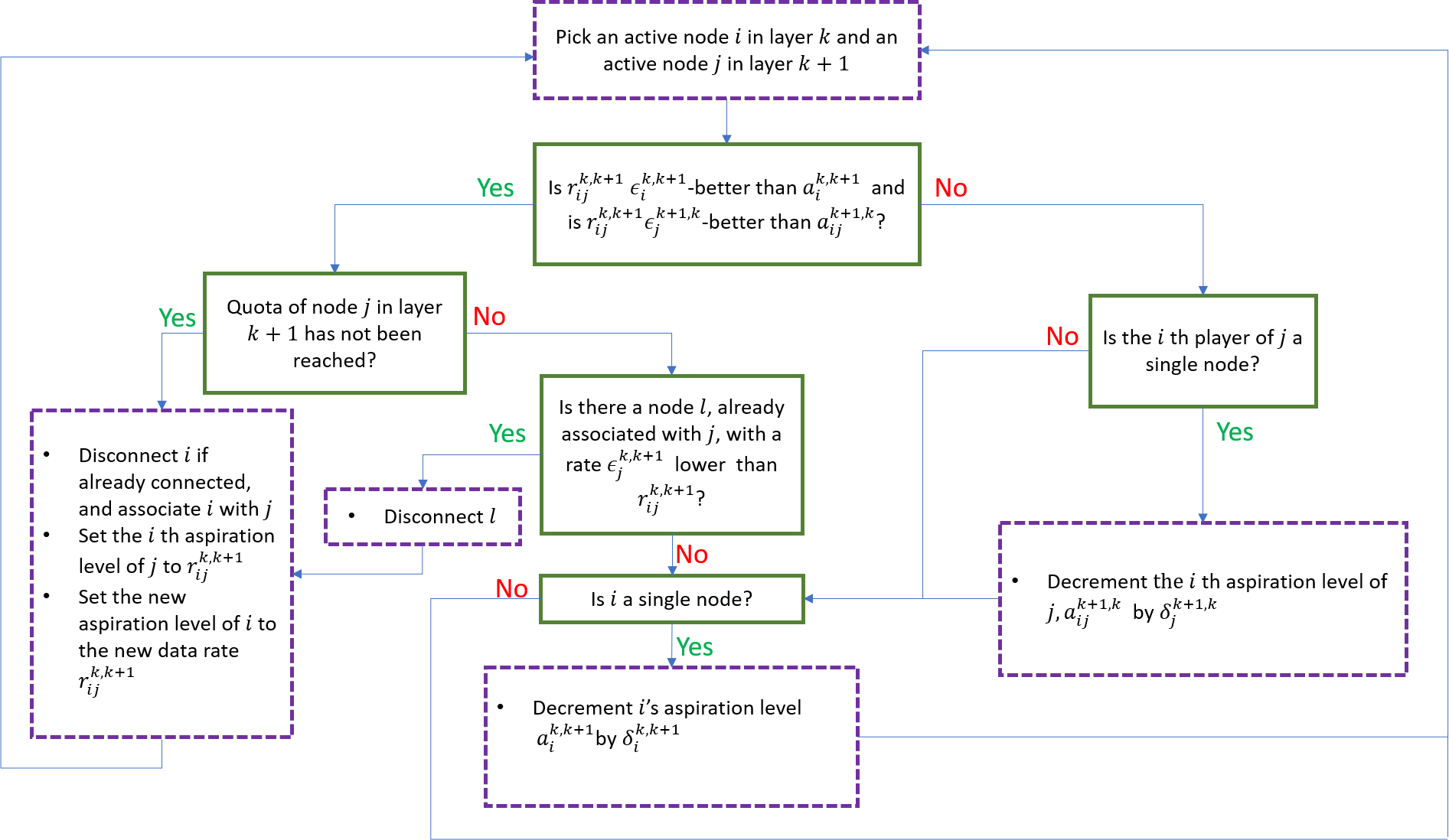}
    \caption{Blind matching algorithm between two adjacent layers $k$ and $k+1$.}
    \label{BlindMatch}
\end{figure*}
A number of works in the mutli-sided assignment literature highlight conditions for non-emptiness of the core, e.g., \cite{atay2017essays, stuart1997supplier,quint1991core}, however very few works construct algorithms to reach them. In our work, we make use of the Blind Matching Algorithm (BLMA) proposed in \cite{hamza2017blma} to reach an efficient matching. The algorithm uses a simple aspiration learning dynamic.  Agents maintain aspiration levels, rates they currently have or have acquired in the past. Agents start out looking for a small improvement, $\epsilon$ over their current aspiration levels. They randomly meet agents upstream or downstream. If both agents can achieve at least an $\epsilon$-improvement, they match and increase their aspiration levels to equal the total value that their pair can generate. Otherwise, if it not possible for the pair to \emph{simultaneously} improve their aspiration levels, and if they are already single, then they know they need to reduce such aspirations. The reduction is done  by another small number $\delta < \epsilon$. This randomized and simple dynamic can be shown to produce an $\epsilon$-pairwise  stable matching \cite{hamza2017blma}.

Fig.~\ref{BlindMatch} presents the blind matching algorithm when implemented between two adjacent layers. First, assume the positive parameters $\epsilon$ and $\delta$ such as $\epsilon>\delta>0$. Suppose
each node initializes its aspiration level at some random value. Then, when an activated node $i$ in layer $k$ encounters an activated upstream node $j$ in layer $k+1$, nodes $i$ and $j$ compare their utilities to their aspiration levels as follows. If the payoffs of both nodes with respect to one another are $\epsilon$-better than their current aspiration levels, and the quota of node $j$ has not been reached yet, node $i$ disconnects from any previous association and associates with node $j$. Otherwise, if node $j$'s quota is reached, aerial platform $j$ checks if the downstream node $i$'s payoff is $\epsilon$-better than one of its served nodes. If so, the aerial platform disconnects the less preferred agent and associates with node $i$. Furthermore, once node $i$ is associated with $j$, it updates its aspiration level with its payoff with respect to $j$. Similarly, node $j$ updates its $i$-th aspiration level with its payoff with respect to $i$. 
In case downstream node $i$ remains single after its interaction with $j$, $i$ reduces its current aspiration level by $\delta$. Moreover, if node $j$ has not reached its quota yet, it also reduces its $i$-th aspiration level by $\delta$. This process is repeated until quotas are reached or all nodes in layer $k$ are connected. 

The blind matching algorithm is guaranteed to achieve the best sum-rate per adjacent layers. It ensures that the number of associated nodes is maximized and also the final aspiration levels of all nodes it returns is a set of payoff in the $\epsilon$-core~\cite{hamza2017blma}. Moreover, the blind matching approach does not require a large amount of information exchange between pairs. Each agent has only to observe its payoff with respect to the target node in the adjacent layer. 

Additionally, it is well worth noting that the blind matching algorithm can allow each node to set a required data rate. To align with these data rate requirements, the first method consists of considering these data rate requirements of ground users as constraints of the maximization problem. Then, during BLMA execution, it should only allow the formation of pairs that respect the data rate requirement of the user. The second method consists of offering the possibility to each node to set its initial aspiration level according to its requirement. The execution of the BLMA may help achieve a data rate close to the initial node's aspiration.

From a complexity perspective, the best bound currently obtained for the BLMA execution between two adjacent layers $\mathcal{S}^k$ and $\mathcal{S}^{k+1}$, $k \in \{0,1,2\}$  was in [18] with a complexity of $O(N_{\text{max}}^3.\max(N_{\text{max}},\frac{\omega^*}{{\delta}}))$ , where $N_{\text{max}}=\max(|\mathcal{S}^k|,|\mathcal{S}^{k+1}|)$  and $\omega^*=\max(\frac{r_{ij}^{k,k+1}}{q_j})_{i\in \mathcal{S}^k,j\in \mathcal{S}^{k+1}}$ is the maximum data rate allocated. By choosing $\delta$, such that $N_{\text{max}}\geq \frac{\omega^*}{{\delta}}$, the convergence to the core will be in $O(N_{\text{max}}^4)$. In practice, as shown through our experimental results, this theoretical bound is not  reached.

As it will be shown in section. V, the BLMA exhibits quick convergence time, enabling multiple executions of the algorithm across various small time intervals with different network configurations.

\section{Performance Evaluation}\label{Performance}


\begin{figure*}
\centering
\begin{tabular}{ccc}
\includegraphics[scale=0.45]{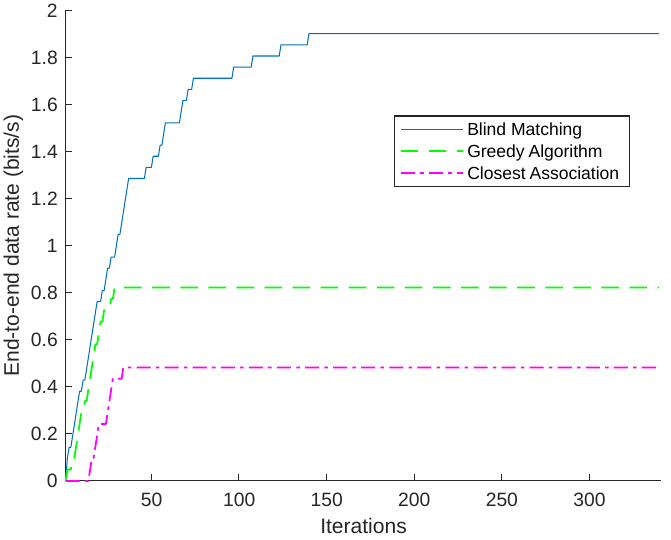}&
\includegraphics[scale=0.45]{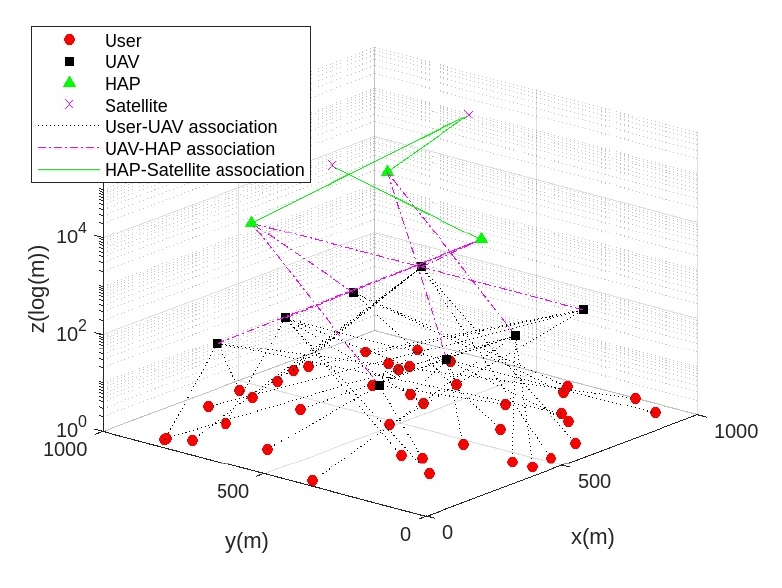}&
\includegraphics[scale=0.45]{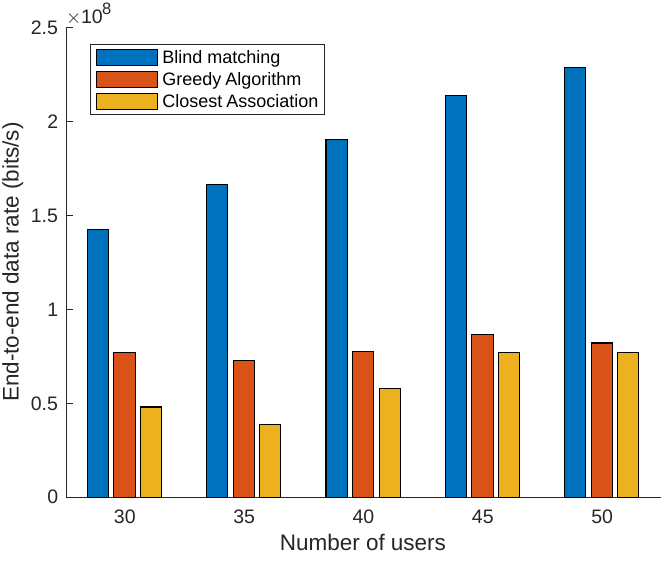}\\
\small{(a)}&\small{(b)}& \small{(c)} 
\end{tabular}
\caption{ {{(a) End-to-end-rate vs iterations, (b) Final association, (c) End-to-end-rate vs number of users}} }
\label{SumRate}
\end{figure*}

\begin{figure}[h]
    \centering
    \includegraphics[scale=0.40]{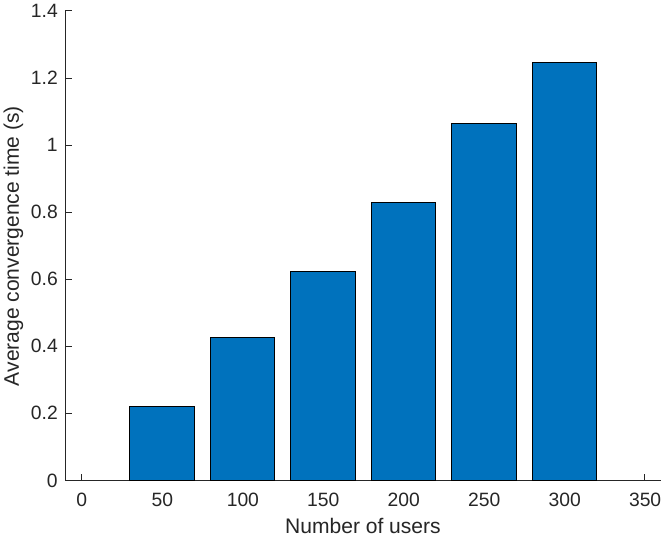}
    \caption{{Average convergence time vs number of users}}
    \label{fig:ConvTime}
\end{figure}

\subsection{Simulation setup}
 We consider a $1\times 1$ $ km^2$ area where $40$ ground devices are randomly scattered. The transmit power of each device is $1$ Watt. We assume $8$ UAVs that hover at $100 m$ of altitude. To transfer the data, UAVs communicate with $3$ HAPs with a power of $3$ Watt. The HAPs are assumed at an altitude of $17 km$. They communicate with two satellites at an orbit height of $700 km$. The transmit power of the HAPs is set to $10$ Watt. We also assume that the HAPs antennas' gain is $45$ dbi, and the noise spectral density $N_0$ is $-169$ dbm/Hz. To model the ground-to-air channel, we assume a suburban environment and we consider the same environment values in~\cite{shi2020joint}. We assume that a node $i$ in layer $k\in\{0,1,2\}$ operates at frequency $f_{c}(k)+B_0^{k,k+1} \times i$ with nodes in layer $k+1$, with
 $f_{c}(k)$ the lowest carrier frequency in layer $k$ and $B_0^{k,k+1}$ the bandwidth over wich layer $k$ nodes communicate with layer $k+1$ nodes.
 The lowest carrier frequencies are $f_{c}(0)=2.5$ GHz, $f_{c}(1)=5$ GHz and $f_{c}(2)=3$ GHz and the bandwidths are $B_0^{0,1}=10^4$ Hz, $B_0^{1,2}=10^5$ Hz and $B_0^{2,3}=10^6$ Hz. We also assume the same quota of $6$ for all UAVs, $3$ for HAPs, and $2$ for satellites.
 
 Experiments were run with an Intel Core i5-7300U CPU running at frequency 2.60 GHz x 2 cores using 16 GB of RAM.
 
To assess the performance of the proposed approach, we consider two other association schemes: \textit{the greedy approach} and \textit{the distance-based association} scheme. The greedy approach is a centralized algorithm whereby a centralized entity first selects the maximum utility of nodes in layer $k$ with respect to nodes in upper layer $k+1$, and associates the corresponding pair. Consequently, the quota of the $(k+1)$-layer node is decremented and the agent in layer $k$ is discarded from the list of unconnected agents. Then, the second
highest utility, within the same layers, is considered, and an association is established between the corresponding pair. This process is repeated until all agents in layer $k$ are
associated or the quota of nodes in layer $k+1$ is reached. Once the associations between adjacent $k$ and $k+1$ tiers are established, nodes in layer $k+1$ calculate their utilities accordingly, and greedily associate the nodes of new adjacent layers. This process starts first from the ground tier and moves up to the space tier.

On the other hand, \textit{the distance-based association}, as the name suggests, assumes that each node in layer $k$ sends a request for an association to its closest node in upper layer $k+1$. When this node receives all the association requests, it selects the agents with the best rates up to its quota. The rejected $k$-layer nodes remain unconnected.

\subsection{Simulation results}

Fig.~\ref{SumRate}(a) plots the end-to-end-rate versus the number of iterations. As depicted in the figure, the blind multi-sided matching approach outperforms both greedy and distance-based association schemes. In fact, the greedy approach selects the best pair of nodes at each iteration. Once the nodes are associated, their association is never revised. Contrarily, the blind matching allows a pair of nodes to disconnect and connect with other nodes whenever an $\epsilon$ improvement is possible. This can significantly improve the end-to-end rate in some scenarios. Furthermore, both greedy and distance-based association schemes converge quickly than the BLMA. However, even at the moment of convergence of greedy and distance-based association algorithms the end-to-end data rate achieved by the BLMA is far superior. For example, at iteration $36$ where the distance-based association algorithm converges, the the end-to-end data rate of the distanced-based association algorithm is $48.31$ Mbits/s, the greedy algorithm's end-to-end data rata is $82.26 $ Mbits/s and the BLMA achieves a total of $123.92$ Mbits/s. By allowing the BLMA to run until it's convergence (in our simulation, we considered it as $300$ successive iterations without data rate improvement), it is possible to obtain better data rate. In fact the BLMA converges after $142$ iterations with an end-to-end data rate of $190.33$ Mbits/s.

In Fig.~\ref{SumRate}(b), we plot the final association between nodes in the 4 layers after the blind multi-sided matching convergence. As it can be seen from the figure, the quota of upper-layer nodes is respected. We can also notice that all ground users are connected, through relays in intermediate layers, to the space tier. This is because the sum of quotas of agents in layer $k+1$ is larger than the number of nodes in layer $k$. 

Fig.~\ref{SumRate}(c) shows the performance of the studied approach for increasing size of the network. The figure plots the final end-to-end rate value against the number of users. As shown in the figure, the blind matching approach always outperforms greedy and closest node associations. In fact, the end-to-end-rate increases with the number of devices for both greedy and blind matching algorithms until the sum of quotas is reached. 


In Fig.~\ref{fig:ConvTime}, we plot the average convergence time of the Blind matching algorithm for different numbers of users. The average convergence time is obtained for a fixed number of device by averaging the convergence time of $10$ network states where nodes are randomly positioned. As illustrated, the execution time increases with the number of users, and the convergence is fast, and happens in average in less than $1.4$ seconds for $300$ users .

\section{Conclusion}
In this paper, we studied the problem of multi-tier association in space-air-ground integrated networks. The multi-layer association is formulated as a multi-sided assignment game. To ensure stable associations across layers and improve the sum of end-to-end rates, we adopted a fully distributed, asynchronous and low-computational algorithm, referred to as the blind multi-sided matching algorithm. Our simulation results showed that significant performance is achieved when compared with the greedy algorithm and distance-based association. 

\section*{Acknowledgment}
This document has been produced with the financial assistance of the European Union (Grant no. DCI-PANAF/2020/420-028), through the African Research Initiative for Scientific Excellence (ARISE), pilot programme. ARISE is implemented by the African Academy of Sciences with support from the European Commission and the African Union Commission. The contents of this document are the sole responsibility of the author(s) and can under no circumstances be regarded as reflecting the position of the European Union, the African Academy of Sciences, and the African Union Commission.


\balance

\bibliographystyle{IEEEbib}
\bibliography{BiblioUAV}

\begin{thebibliography}{10}

\bibitem{zhang2023HAP}
Y.~Zhang, M.A. Kishk, and M.-S. Alouini,
\newblock ``Hap-enabled communications in rural areas: When diverse services
  meet inadequate communication infrastructures,''
\newblock {\em IEEE Open Journal of the Communications Society}, vol. 4, pp.
  2274--2285, 2023.

\bibitem{ndiaye2022age}
M.~N. Ndiaye, E.~Bergou, M.~Ghogho, and H.~El Hammouti,
\newblock ``Age-of-updates optimization for {UAV}-assisted networks,''
\newblock in {\em 2022 IEEE Global Communications Conference}. IEEE, 2022, pp.
  450--455.

\bibitem{liu2018space}
J.~Liu, Y.~Shi, Z.~M. Fadlullah, and N.~Kato,
\newblock ``Space-air-ground integrated network: A survey,''
\newblock {\em IEEE Communications Surveys \& Tutorials}, vol. 20, no. 4, pp.
  2714--2741, 2018.

\bibitem{jia2022SAGIN}
Z.~Jia, M.~Sheng, J.~Li, and Z.~Han,
\newblock ``Toward data collection and transmission in 6g space air ground
  integrated networks: Cooperative hap and leo satellite schemes,''
\newblock {\em IEEE Internet of Things Journal}, vol. 9, no. 13, pp.
  10516--10528, 2022.

\bibitem{el2021optimal}
H.~El Hammouti, D.~Hamza, B.~Shihada, M.-S. Alouini, and Jeff J.~S~Shamma,
\newblock ``The optimal and the greedy: Drone association and positioning
  schemes for internet of {UAVs},''
\newblock {\em IEEE Internet of Things Journal}, 2021.

\bibitem{alsharoa2020improvement}
A.~Alsharoa and M.-S. Alouini,
\newblock ``Improvement of the global connectivity using integrated
  satellite-airborne-terrestrial networks with resource optimization,''
\newblock {\em IEEE Transactions on Wireless Communications}, vol. 19, no. 8,
  pp. 5088--5100, 2020.

\bibitem{abderrahim2020latency}
W.~Abderrahim, O.~Amin, M.-S. Alouini, and B.-Shihada,
\newblock ``Latency-aware offloading in integrated satellite terrestrial
  networks,''
\newblock {\em IEEE Open Journal of the Communications Society}, vol. 1, pp.
  490--500, 2020.

\bibitem{shi2020joint}
Y.~Shi, Y.~Xia, and Y.~Gao,
\newblock ``Joint gateway selection and resource allocation for cross-tier
  communication in space-air-ground integrated {IoT} networks,''
\newblock {\em IEEE Access}, vol. 9, pp. 4303--4314, 2020.

\bibitem{arabi2019rat}
S.~Arabi, H.~El Hammouti, E.~Sabir, H.~Elbiaze, and M.~Sadik,
\newblock ``{RAT} association for autonomic {IoT} systems,''
\newblock {\em IEEE Network}, vol. 33, no. 6, pp. 116--123, 2019.

\bibitem{gale1962college}
D.~Gale and L.~S Shapley,
\newblock ``College admissions and the stability of marriage,''
\newblock {\em The American Mathematical Monthly}, vol. 69, no. 1, pp. 9--15,
  1962.

\bibitem{roth1992two}
A.~E Roth and M.~Sotomayor,
\newblock ``Two-sided matching,''
\newblock {\em Handbook of game theory with economic applications}, vol. 1, pp.
  485--541, 1992.

\bibitem{shapley1971assignment}
Lloyd~S Shapley and Martin Shubik,
\newblock ``The assignment game i: The core,''
\newblock {\em International Journal of game theory}, vol. 1, no. 1, pp.
  111--130, 1971.

\bibitem{hamza2017blma}
D.~Hamza and J.~S Shamma,
\newblock ``{BLMA}: A blind matching algorithm with application to cognitive
  radio networks,''
\newblock {\em IEEE Journal on Selected Areas in Communications}, vol. 35, no.
  2, pp. 302--316, 2017.

\bibitem{atay2017essays}
A.~Atay,
\newblock ``Essays on multi-sided assignment markets,''
\newblock 2017.

\bibitem{zhao2018resource}
J.~Zhao, Y.~Liu, T.~Mahmoodi, K.~K. Chai, Y.~Chen, and Z.~Han,
\newblock ``Resource allocation in cache-enabled {CRAN} with non-orthogonal
  multiple access,''
\newblock in {\em 2018 IEEE International Conference on Communications (ICC)},
  2018, pp. 1--6.

\bibitem{liu2019joint}
J.~Liu, G.~Wu, S.~Xiao, X.~Zhou, G.~Y. Li, S.~Guo, and S.~Li,
\newblock ``Joint power allocation and user scheduling for
  device-to-device-enabled heterogeneous networks with non-orthogonal multiple
  access,''
\newblock {\em IEEE Access}, vol. 7, pp. 62657--62671, 2019.

\bibitem{hammouti2018air}
H.~El Hammouti and M.~Ghogho,
\newblock ``Air-to-ground channel modeling for {UAV} communications using {3D}
  building footprints,''
\newblock in {\em Ubiquitous Networking, UNet 2018, Revised Selected Papers 4}.
  Springer, 2018, pp. 372--383.

\bibitem{stuart1997supplier}
H.~W~Stuart Jr,
\newblock ``The supplier--firm--buyer game and its m-sided generalization,''
\newblock {\em Mathematical Social Sciences}, vol. 34, no. 1, pp. 21--27, 1997.

\bibitem{atay2018bargaining}
A.~Atay and T.~Solymosi,
\newblock ``On bargaining sets of supplier-firm-buyer games,''
\newblock {\em Economics Letters}, vol. 167, pp. 99--103, 2018.

\bibitem{quint1991core}
T.~Quint,
\newblock ``The core of an m-sided assignment game,''
\newblock {\em Games and Economic Behavior}, vol. 3, no. 4, pp. 487--503, 1991.

\end{thebibliography}
\end{document}